\documentclass[sigconf]{acmart}

\AtBeginDocument{%
  }

\newcommand{\revise}[1]{#1}

\usepackage{multirow}

\usepackage{xcolor,tikz}

\usepackage{acmart-taps}

\newcommand*\encircled[2][000000]{%
  \begingroup
  \definecolor{enc@fill}{HTML}{#1}
  \tikz[baseline={([yshift=-.45ex]char.base)}]{%
    \node[
      circle,
      fill=enc@fill,      
      draw=enc@fill,
      text=white,
      font=\scriptsize\bfseries,
      inner sep=0.08em,
      minimum size=1.0em,
      text height=1ex,
      text depth=0pt
    ] (char) {#2};
  }%
  \endgroup
}
\makeatother

\copyrightyear{2026}
\acmYear{2026}
\setcopyright{cc}
\setcctype{by}
\acmConference[CHI '26]{Proceedings of the 2026 CHI Conference on Human Factors in Computing Systems}{April 13--17, 2026}{Barcelona, Spain}
\acmBooktitle{Proceedings of the 2026 CHI Conference on Human Factors in Computing Systems (CHI '26), April 13--17, 2026, Barcelona, Spain}
\acmPrice{}
\acmDOI{10.1145/3772318.3790425}
\acmISBN{979-8-4007-2278-3/2026/04}

\acmSubmissionID{4143}

\begin{document}


\title{Robot-Assisted Group Tours for Blind People} 
\def\etal{{\it et al.}}

\settopmatter{authorsperrow=4}
\author{Yaxin Hu}
\orcid{0000-0003-4462-0140}
\affiliation{%
  \institution{Department of Computer Sciences\\University of Wisconsin--Madison}
  \streetaddress{Department of Computer Sciences, University of Wisconsin--Madison}
  \city{Madison}
  \state{Wisconsin}
  \country{USA}
}
\email{yaxin.hu@wisc.edu}

\author{Masaki Kuribayashi}
\orcid{0000-0001-8412-223X}
\affiliation{
  \institution{Miraikan - The National Museum of Emerging Science and Innovation}
  \city{}
  \country{}
}
\affiliation{
  \institution{Waseda University}
\city{Tokyo}
  \country{Japan}
}
\email{rugbykuribayashi@toki.waseda.jp}

\author{Allan Wang}
\orcid{0000-0002-8253-2742}
\affiliation{
  \institution{Miraikan - The National Museum of Emerging Science and Innovation}
  \city{Tokyo}
  \country{Japan}
}
\email{allan.wang@jst.go.jp}

\author{Seita Kayukawa}
\orcid{0000-0002-0678-1157}
\affiliation{
  \institution{IBM Research - Tokyo}
  \city{Tokyo}
  \country{Japan}
}
\email{seita.kayukawa@ibm.com}

\author{Daisuke Sato}
\orcid{0000-0002-7670-9177}
\affiliation{
  \institution{Robotics Institute \\ Carnegie Mellon University}
  \city{Pittsburgh}
  \state{Pennsylvania}
  \country{USA}
}
\email{daisukes@cs.cmu.edu}

\author{Bilge Mutlu}
\orcid{0000-0002-9456-1495}
\affiliation{%
  \institution{Department of Computer Sciences\\University of Wisconsin--Madison}
  \streetaddress{Department of Computer Sciences, University of Wisconsin--Madison}
  \city{Madison}
  \state{Wisconsin}
  \country{USA}
}
\email{bilge@cs.wisc.edu}

\author{Hironobu Takagi}
\orcid{0000-0003-3087-3251}
\affiliation{
  \institution{IBM Research - Tokyo}
  \city{Tokyo}
  \country{Japan}
}
\email{takagih@jp.ibm.com}

\author{Chieko Asakawa}
\orcid{0000-0002-5447-1305}
\affiliation{
  \institution{IBM Research\\IBM}
  \city{Yorktown Heights}
  \state{New York}
  \country{USA}
}
\affiliation{
  \institution{Miraikan - The National Museum of Emerging Science and Innovation}
  \city{Tokyo}
  \country{Japan}
}
\email{chiekoa@us.ibm.com}

\renewcommand{\shortauthors}{Hu et al.}


\begin{abstract}
Group interactions are essential to social functioning, yet effective engagement relies on the ability to recognize and interpret visual cues, making such engagement a significant challenge for blind people. In this paper, we investigate how a mobile robot can support group interactions for blind people. We used the scenario of a guided tour with mixed-visual groups involving blind and sighted visitors. Based on insights from an interview study with blind people ($n=5$) and museum experts ($n=5$), we designed and prototyped a robotic system that supported blind visitors to join group tours. We conducted a field study in a science museum where each blind participant ($n=8$) joined a group tour with one guide and two sighted participants ($n=8$). Findings indicated users' sense of safety from the robot's navigational support, concerns in the group participation, and preferences for obtaining environmental information. We present design implications for future robotic systems to support blind people's mixed-visual group participation.

\end{abstract}

\begin{teaserfigure}
  \centering
  \includegraphics[width=0.93\textwidth]{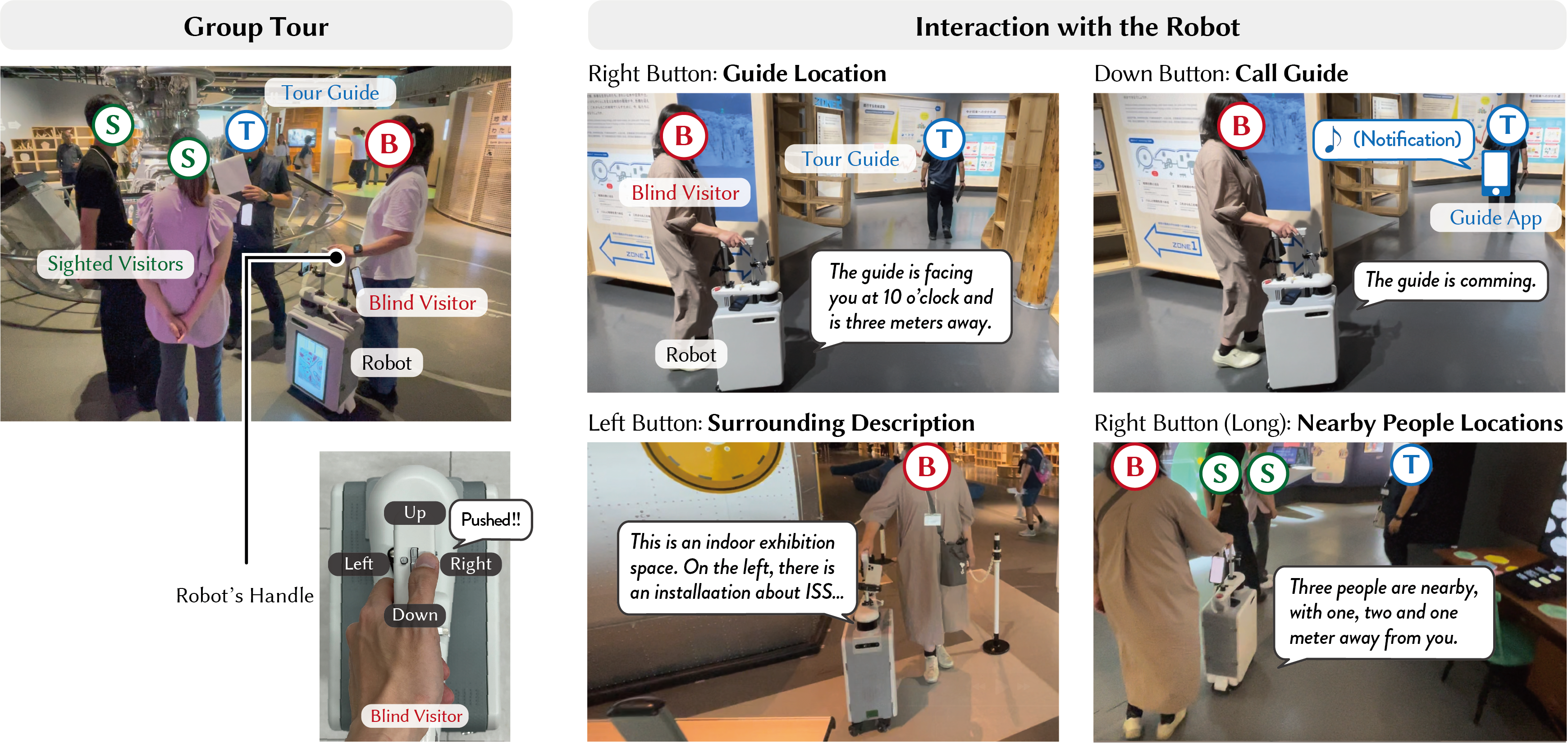}
  \vspace{-10pt}
  \caption{In this paper, we investigated how assistive robots can support blind people to participate in group tours with sighted people. We conducted a three-phase study involving an \textit{Interview Study}, \textit{Robot Design \& System Building}, and a \textit{Field Study}. Our findings highlighted the needs and challenges for blind people's tour engagement and interaction patterns with the assistive robot. \textit{\textbf{Left}}: One blind visitor joined a group tour with two sighted visitors and one tour guide. The blind user can press the four buttons on the handle of the robot to trigger assistive features. \textbf{\textit{Right}}: Four robotic features supporting group tour participation. }
  \label{fig:teaser}
  \Description{Teaser figure with left and right sides.  
LEFT side: there are two pictures on the left side. 
One picture shows a blind user participating in a group tour with the support of the robot in a science museum.  
Another picture shows a handle with four buttons on the robot labelled as "UP", "DOWN", "LEFT" and "RIGHT". During the tour, the blind user can press the four buttons on the handle of the robot to trigger different types of features. 
RIGHT side:  four pictures were displayed demonstrating robot features. The four pictures are described below: 
Top-left: Requesting the position of the guide. One blind person is holding a robot with the shape of suitcase. A tour guide is standing at some distance from the blind person. A speech button has text: The guide is facing you at 10 o'clock and is three meters away. 
Top-right: Requesting assistance from the guide. One blind person is holding a robot with the a shape of suitcase. A tour guide is standing at some distance from the blind person.  A speech button has texts: The guide is coming. 
Bottom-left: Describe the surrounding environment and exhibits. One blind person is holding a robot with a shape of a  suitcase standing next to an exhibition about international space station (ISS).  A speech button has texts: This is an indoor exhibition space. On the left, there is an installation about ISS. 
Bottom-right: Requesting quantity and distance of surrounding people. Four people are in the picture. One blind person is holding a robot with a shape of a  suitcase.  Two sighted visitors are standing in the exhibition space. A tour guide is also standing in the exhibition space. A speech button has texts: Three people are nearby, with one, two and one meter away from you.}
\end{teaserfigure}
\maketitle

\section{Introduction}
Assistive robots have been widely studied to support blind people's everyday activities, including indoor and outdoor navigation \cite{azenkot2016enabling, guerreiro2019cabot}, physical assembly tasks \cite{bonani2018my}, and inclusive classroom activities \cite{neto2021community, neto2023robot}. These efforts typically emphasize capabilities such as perceiving the environment, planning and executing tasks, and conversing with their users, all in service of supporting independence in daily living. Far less is known about how assistive robots can support blind people in navigating \textit{social} and \textit{group} settings. 

A group involves two or more individuals who are connected by social relationships \cite{forsyth2014group}. Participation in groups offers new information and resources and creates shared social experiences for each member in the group. However, successful group interactions require the ability to interpret visual information of the group and the surroundings, and take actions according to group dynamics, which can cause significant challenges for blind people \cite{naraine2011social}. Assistive technologies have been studied to \revise{promote independence and autonomy} for blind people to participate in group conversations \cite{phillips2018social} and collaborative tasks \cite{mendes2020collaborative}. Common technical platforms include wearable devices \cite{ye2014current}, collaborative tabletops \cite{mendes2020collaborative}, haptic devices \cite{phillips2018social}, and smart glasses \cite{krishna2008systematic}. However, existing approaches primarily focus on visual and informational support for group activities, which is not sufficient to support blind people in \revise{taking actions in physical group activities independently in dynamic and complex environments}.

In this paper, we investigate how assistive mobile robots can support blind visitors to participate in group activities \revise{independently} in a dynamic public space. We focus on a guided group tour scenario in a science museum and designed and prototyped an assistive robot to support blind people in participating in tours with sighted people. We specifically pose the following \revise{three} research questions: 

\begin{itemize}
    \item $RQ_1.$ \revise{What are blind people's needs and challenges in group participation?} 
    \item $RQ_2.$ How can we support these needs through the design of a mobile robotic assistant?
    \item $RQ_3.$ How would robot-assisted participation affect other members in the group?
\end{itemize}

We followed a three-phase research process to answer these questions. First, we conducted an interview study with blind people ($n=5$) and museum experts ($n=5$) in a science museum to understand blind people's needs and challenges in group tours. Second, insights from the first study guided us to design a robot with features to (1) help the blind user follow the guide, (2) help the blind user signal to the guide, (3) describe the positions of the guide and nearby people, and (4) describe the surrounding environment to the blind user. Third, we conducted a field study with blind participants ($n=8$) and sighted participants ($n=8$) where each tour group had one guide, one blind participant, and two sighted participants. 

Our findings from the interview study revealed how tours for blind people were designed and conducted, challenges faced by the tour guide and blind people, and robot design insights for supporting mixed visual group tours. These findings led to three design goals for implementing the robotic system: (1) communication support, (2) environmental awareness, and (3) independence in navigation. Our field study revealed patterns in which the robotic system affected blind participants' ability to follow and engage with the group during the tour. Findings emphasized the importance of providing users with a sense of safety; supporting user control and effective robot feedback during navigation; balancing information from the robot and the guide; and helping users maintain connections with the guide on the tour.  

Our contributions are as follows:

\begin{itemize}
    \item \textit{Design requirements:} Through an interview study with blind people and museum experts, we developed an understanding of the structure and challenges of conducting tours for blind people and requirements for the robotic system design. 
    \item \textit{System implementation:} Building on this understanding and requirements, we designed and implemented a robotic system to support group tours for blind people.
    \item \textit{Empirical findings:} Our field study revealed robot use patterns, particularly the ways in which the robot assisted group following and tour engagement for blind people in mixed visual tour groups.  
\end{itemize}

\section{Related Work}
\subsection{Assistive Robots for Blind and Visually Impaired People}
Navigation robots have recently gained increasing attention for guiding blind people~\cite{wei2025human,kulkarni2016robotic,stals2025robot}. 
Their appeal lies in their ability to provide automatic mobility and wayfinding~\cite{albogamy2021sravip,zhang2019human,wei2025human}, typically through real-time sensing, infrastructure support, maps, and localization methods~\cite{bineeth2020blindsurvey,sulaiman2021analysis}. 
A variety of robotic forms have been explored, including drones~\cite{liao2023running,huppert2021guidecopter}, companion-like~\cite{feng2015designing}, cane-like~\cite{ranganeni2023exploring}, quadruped~\cite{zhang2023follower,wang2021navdog,kim2023transforming}, and wheeled designs~\cite{lu2021assistive,zhang2023follower,guerreiro2019cabot}. 
Each has distinct advantages and limitations: for example, quadruped robots can manage stairs and uneven outdoor terrains~\cite{cai2024navigating}, yet wheeled robots are still generally preferred because they are quieter, more stable, and smoother in motion~\cite{wang2022can, hu2024really, hu2025narraguide}. Since robots can take over mobility and wayfinding, blind users are able to focus more on perceiving their surroundings~\cite{kayukawa2022HowUsers}. 
This capability has extended the applicability of navigation robots beyond point-to-point travel~\cite{wei2025human} to tasks such as running~\cite{liao2023running}, shopping mall exploration~\cite{kuribayashi2025wanderguide}, and interactive navigation in unfamiliar buildings~\cite{kuribayashi2023pathfinder}.
\revise{One of the most relevant works is the autonomous museum guide robot by Kayukawa~\etal~\cite{kayukawa2023enhancing}. The robot allows users to select exhibitions of interest, navigate to the entrance of the exhibitions, and request staff assistance upon arrival. With dedicated support from museum staff, users can explore the exhibits in detail. 
Here we study the context of group tours, which include multiple stakeholders such as blind users, other tour participants, and tour guides. 
As discussed in a later section, the group tour context requires unique interaction designs beyond point-to-point navigation and requesting support. It needs to support interpersonal relationships, including sharing guide information with users, communicating user-specific needs to guides, coordinating participant details, and supporting tour group following. Since these requirements remain underexplored, we reveal them through interviews with each stakeholder in the group tour.}




\subsection{Group Activities for Blind People}
\label{sec:group-blind-people}
Group interactions are essential in daily lives, benefit individuals in social connections and support instrumental goals \cite{forsyth2014group, mcgrath1991time}. Existing work discusses how robots can facilitate group activities~\cite{short2017robot, vazquez2017towards, hu2025designing}. However, group interactions pose significant challenges for blind people due to the loss of visual cues in engaging in groups. Prior work has investigated blind people's interaction with sighted people in a variety of group contexts, such as social communications \cite{jones2024don, jones2025put, phillips2018social}, collaborative tasks \cite{mendes2020collaborative, branham2015collaborative}, and social virtual reality (VR) games \cite{collins2023guide, ji2022vrbubble}. Specifically, \citet{jones2024don} studied the joint attention challenges faced by blind people and revealed their concerns for turn-taking and voids in group communications. In the same vein, prior work has proposed the use of assistive technologies to support blind people's group interactions, through describing the social environment \cite{yaagoubi2008cognitive} and aiding in conversational signaling \cite{dyzel2020assistive, freitas2008speech}. Common platforms included wearable devices \cite{ye2014current}, tabletops \cite{mendes2020collaborative}, haptic devices \cite{phillips2018social}, and smart glasses \cite{krishna2008systematic}. Different from wearable and mobile devices, robots can provide both navigational and information support to assist blind people in group interactions in dynamic and complex environments. In our work, we focused on the scenario of guided group tours and study how robots can assist blind people to join the tour with sighted visitors.

\subsection{Guided Tours for Blind People}
Leisure experience is an important part of our life, but blind people often face many barriers to participating in recreational activities \cite{bandukda2020places, li2023understanding}. One important way of engaging in leisure activities is through guided tours, which are curated with experts' knowledge to enhance visitors' experience. Prior work has studied the need and design of human-guided tours for blind people in a variety of leisure contexts, such as art galleries \cite{kreplak2014artworks}, gardens \cite{wang2024translating}, and theaters \cite{udo2010enhancing}. In addition, assistive mobile robots were commonly used as tour guides for blind people \cite{kayukawa2023enhancing, asakawa2019independent} and supported them to follow tour routes and provide exhibit information. However, robot-guided tours often target individual visitors \cite{hu2025narraguide} and lack social interactions and personalized storytelling unique in human guided tours. Although many cultural institutions provide accessible tours for blind visitors, these special tours occur much less frequently compared to regular tours, which are often not accessible to blind visitors. To this end, we aim to understand how robots can help blind people participate in mixed-visual tour groups guided by human guides and enhance independence in their leisure and social engagements.

\begin{figure*}
    \centering
    \includegraphics[width=\linewidth]{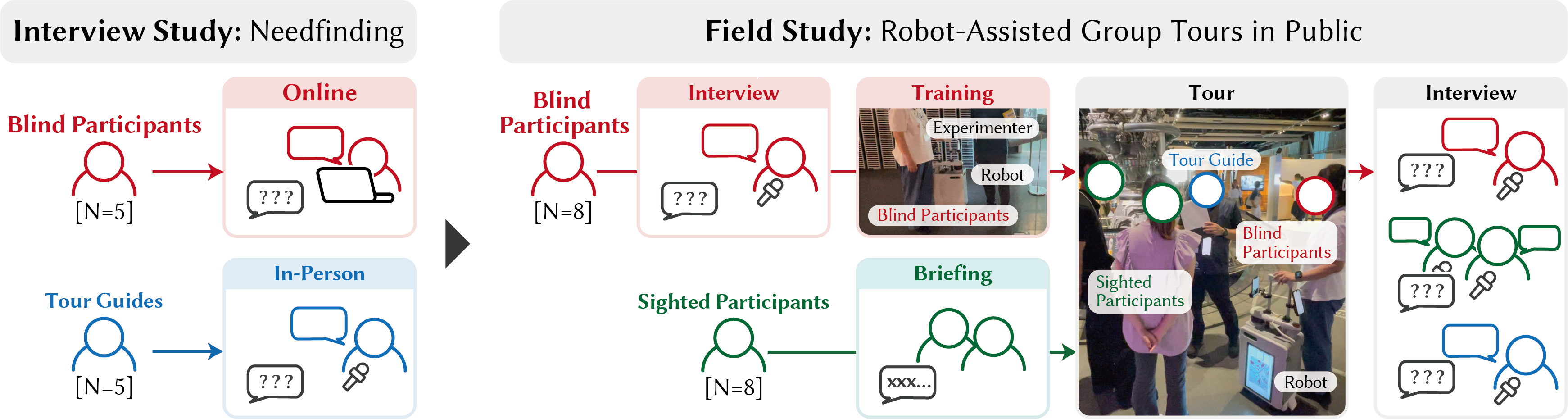}
    \caption{\textbf{Study process overview.} Left: We first conducted interviews with both blind participants and museum experts, \textit{i.e.}, science communicators (SCs), who have experience with accessibility projects. Based on the reported challenges and opportunities, we developed a robotic system to aid group tour participation. Right: We evaluated the tour experience with the robot by recruiting blind participants, sighted tour visitors, and a tour guide. For each blind participant, a training session was held with the guide before the tour.}
    \Description{Study process overview.
 LEFT: Study process for Interview Study: Needfinding 
Icons showing interviews with five blind participants and five tour guides 

RIGHT: Field Study process: Robot-Assisted Group Tours in Public 
Icons showing interviews with blind participants (N=8) and sighted participants (N=8)
One picture labelled as "Training" showing training session where a blind participant is trying the robot with the companion of an experimenter 
One picture labelled as "Tour" showing one blind participant with a robot, two sighted participants and one tour guide.
A set of icons showing interviews with blind participants, sighted participants and tour guides. 
}
    \label{fig:study process}
\end{figure*}

\section{Interview Study: Needfinding}
We first conducted an interview study to understand the design, needs, and challenges of group tours from both blind people's and tour guides' perspectives. All study protocols and materials in this paper were approved by Waseda University's institutional review board (IRB). 

\subsection{Participants}
We advertised our experiments to the local blind and low-vision community via a sizable mailing list maintained by our institution. Using this method, we recruited five blind people. All participants were \revise{legally blind} and were between 49--79 years old ($M=57.00, SD=11.19, Women=4, Men=1$). Four of the participants have participated in guided group tours in a museum. One has participated in a lecture-style group activity. \revise{All five participants have used similar robots in a suitcase shape that supported independent navigation}. 


We further recruited five experts from a science museum through the museum's management team. These experts are the museum's science communicators (SCs), who are in charge of explaining exhibitions to visitors and conducting tours. SCs were 28--32 years old ($M=29.20, SD=1.60, Women=2, Men=3$) and have worked in the museum for 1.5--4 years ($M=2.4, SD=0.97$). Three of them have experience in hosting tours or events for blind and low-vision people, and the other two have experience in organizing tours for the deaf and hard-of-hearing community.

\subsection{Study Procedures}

\paragraph{Interviews with Blind People}
We conducted semi-structured interviews with blind participants online through a teleconferencing tool\footnote{Zoom: \url{https://www.zoom.com/}}. The interview questions were structured into two parts. The first part asked blind people about their past tour experience and how they interacted with other people on the tour. The second part included questions related to blind participants' previous experience with navigational support robots, and how such robots might support them to join museum group tours if they have used the robot before. The interviews lasted 30 minutes. 

\paragraph{Interview with Science Communicators (SCs)}
The study sessions with science communicators (SCs) were conducted in a science museum conference room. The museum provides two types of tours: regular tours for all visitors and special tours dedicated to blind people. We first interviewed SCs about their experience in conducting both regular tours and blind people tours. Interview questions included the tour process, guide strategies, interactions among the tour participants, and differences between sighted visitor tours and blind visitor tours. 

After the first part of the interview questions, SC participants watched a two-minute video clip demonstrating an assistive mobile robot in a suitcase shape and its navigational and conversational features. Although all SCs are familiar with the robot, as it is already part of the museum exhibition, we used the video to help them recall the capabilities of the robot. In the video, a blind user used the robot to navigate in a shopping mall and ask questions about the nearby environment. After watching the video, participants generated ideas on how the robot can be used to support blind visitors in joining group tours. The study session with SCs lasted for one hour. All interviews from the study were audio recorded for data analysis.



\subsection{Data Analysis} \label{formative-analysis}
We conducted thematic analysis on the interview data of the blind participants and SCs from the interview study \revise{following the principle of reflective thematic analysis~\cite{byrne2022worked, braun2019reflecting}. The lead author of the paper first read through the entire dataset thoroughly. Then the lead author independently coded a subset of the interview data and generated an initial codebook. These codes were reviewed and verified by a second coder from the research team. Any disagreements between the two coders were resolved through discussion. With the initial codebook, the lead authors continued coding the remaining data with the second coder performing verification. If new codes were deemed necessary to be introduced, the codebook would be updated iteratively. After all data were coded, the lead author assembled all codes into initial candidate themes and sub-themes using affinity diagramming. The research team reviewed and discussed these themes and generated a final set of themes that reflect different aspects of the dataset core to the research questions.}

\revise{The thematic analysis was conducted by a team of researchers with backgrounds in human robot interaction and accessibility. The research team included one Asian female Ph.D. researcher, one Asian male Ph.D. researcher, and one Asian male post-doc researcher aged from 25--35 years old. Their personal and cultural backgrounds informed their interpretation of the data. All researchers from the team participated in the interview study sessions. They also engaged in reflective discussions throughout the analysis process to reduce potential biases.}

\subsection{Findings}

\subsubsection{Tour Challenges Faced by Blind Visitors}
Our blind participants reported challenges in participating in group tours. 
\paragraph{\textbf{Tour Challenge 1: communication with the tour guide}}
When asked about interacting with the guide on a tour, participants (P1, P2, P3, P5) reported challenges in approaching the guide, listening to the guide in noisy environments, and finding the appropriate time to ask questions, especially in big groups. One participant (P5) shared the difficulty in approaching the guide when she was at the end of the group line, \textit{``If I had just stuck to the line, I probably wouldn't have been able to talk to the guide.''} Another participant (P1) worried about the social norms in having conversations on a tour and shared an unpleasant experience of being scolded by other visitors on a group tour she joined before, as she reported, \textit{``Once... A helper was explaining something to me, and someone said, `You're being loud, please stop.' I've had that experience''} (P1). In addition, one participant (P5) reported not hearing the guide well in a noisy environment: \textit{``If I happened to be at the back of the line, I couldn't hear well, and I had move on to the next booth without fully understanding''} (P5).

\paragraph{\textbf{Tour Challenge 2: difficulties in following the tour}}
Navigation challenges on tours were frequently mentioned in the interviews. Participants reported the feeling of \textit{``anxiety of moving around''} (P3) and \textit{``get[ting] tired of being extra careful''} with the guide dog (P2) in group tours. For example, P3 worried about the lack of support when moving with a big group, stating, \textit{``Even if they know there are visually impaired people, if there are, say, 10 people on the tour, it's hard for them to keep an eye on everyone. I'm worried about that.''} Another participant (P5) reported an experience of falling behind the tour because of not noticing the group movement: \textit{``When I was at the back of the line, I'd sometimes look at the previous exhibit for too long and be late for the next one.''} (P5).

\paragraph{\textbf{Tour Challenge 3: lack of personalized tour pace}}
Two blind participants (P1, P2) preferred to take their own time to visit the exhibits rather than join group tours since they could move at their own pace if visiting alone. For example, P1 stated that \textit{``I probably wouldn't be able to touch things as much as I'd like''} (P1). Similarly, P2 explained she didn't enjoy group tours because \textit{``I would move on before I could form a mental image. So even though I went to the trouble of going, I didn't get much out of it.''} (P2).

\subsubsection{Group Tour for Blind Visitors}
Our SC participants reported the differences between conducting tours for blind visitors and for sighted visitors, and the challenges they encountered when guiding the tours. 

\paragraph{\textbf{Blind Tour Design 1: focus on the touch experience}}
One major difference between tours for blind people and sighted people was the focus on the touch experience. SCs shared the importance of blind visitors touching the exhibits, saying that \textit{``For them to touch and, you know, they experience the details and the structure of the thing.''} (SC4). Furthermore, satisfying experiences involved careful touch and post-visit discussion sessions, which could require longer tour durations. As SC5 pointed out, \textit{``We have separate sections for touching, moving, experiencing things in the exhibit hall, and coming back for dialogue. So if the time is short, it's difficult to provide a satisfying tour for visually impaired people.''} (SC5).



\paragraph{\textbf{Blind Tour Design 2: explain visual information}}
In addition to the touch experience, SCs emphasized the importance of providing visual descriptions of exhibits and the environment (SC1, SC2, SC5). Exhibit information included exhibits' shape, color, position, distance to the visitor, \textit{etc.} Environmental information included descriptions of the current room and people nearby, such as people's costumes, genders, \textit{etc.} Although tour guide scripts were provided ahead of time, SCs reported challenges in giving visual-based explanations and their worries of not conveying the information effectively (SC1, SC2). For example, SC1 shared the difficulty in explaining the model of the \textit{International Space Station} (ISS), \textit{``The shape of the ISS is sometimes described as dragonflies with their tails stuck together, but they've never seen a dragonfly. Some people can't picture that it has four wings, two pairs, in their head. And I often get feedback that that way of explaining is a bit hard to understand.''} (SC1).

\paragraph{\textbf{Blind Tour Design 3: tour group management}}
SCs also reported the need to have additional museum staff to manage tour flow to ensure safety and support touch activities (SC1, SC5). On the group tour for blind visitors, the guide often asked blind visitors to form a line and touch the objects one by one. The guide would speak in front of the line and a museum staff member would stay at the line's end to manage the group flow or \textit{``the traffic from sighted visitors''} (SC5). Sometimes companions of the blind visitor would help to manage the group. As SC1 reported, \textit{``If there is a companion [for the blind visitor], staff will be in front and behind, and we ask the companion to handle the basic movement. But if they come alone, a staff member will be there to support them, next to them.''} In addition, sometimes one more museum staff would help to hold the touchable model, as SC5 shared, \textit{``If we have them touch something along the way, we need someone to hold that object, so in that way, we end up with a three-person team.''} As a result, the group tours designed for blind visitors would \textit{``need more staff than on a regular tour.''} (SC1).

\subsubsection{Robot Design Insights} 
\revise{We report findings on robot design insights from both blind participants and SCs. Participants proposed rich design ideas on how the robot could provide explanations, support group interactions, and assist in independent navigation. In particular, we highlighted robot design insights on alleviating social burdens of asking questions, identifying unknown objects on the way, and supporting social navigation and group positioning for blind people.}  

\paragraph{\textbf{Robot Design 1: explain visual, tactile and environmental information \& point out unknown things}}
\revise{
Blind participants proposed that the robot could describe visual information of exhibits (P1, P2, P4), \textit{``read the signs''} (P3), provide explanations of the exhibition and room layouts (P1--P5), and point out new and unknown things (P4). In particular, one participant (P4) emphasized the need to describe visual information first in the tour group with sighted people: \textit{``Because for sighted people, that part is a given, so they skip that and go straight to the background of the artwork. But first, I want to have the painting or the exhibit itself explained to me, what it's like.''} (P4). In addition, blind participants (P2, P4) suggested that the robot could alleviate their social burdens of asking questions aloud. As one participant (P4) suggested, \textit{``In a museum, you're not supposed to talk much, and it's also a bit difficult to ask my family or a guide. So, if a robot or something could do that for me, it would be a great help.''} Furthermore, one participant (P4) reported a \textit{``window shopping challenge''} she faced: \textit{``We can only go to places we've already decided on... If there's something I don't know about, I'll just pass by it... I think it's really wonderful to learn about new things I didn't know about, but I can't do that.''} This participant (P4) suggested that the robot could point out new and unknown things for them, saying, \textit{``It could say, `Oh, there's this too.' If it had a sort of `window shopping' function, I think that would be nice.''} (P4).}


SCs also proposed design ideas on how the robot could support the tour guide in conducting tours with blind people. Similar to blind participants' suggestions, they mentioned that the robot could describe the visual information of the object (SC4, SC5), whether the exhibition is touchable (SC1, SC3, SC5), and information that \textit{``the human guided tour doesn't offer''} (SC4) or is not knowledgeable about (SC2, SC3). For example, SC3 described how the robot could guide the blind visitor to touchable exhibits, \textit{``It would be great if the [robot] could also navigate and say things like, `This exhibit can be touched.' Or, `There's this kind of device, so please try experiencing it.'''} In addition, one participant (SC2) proposed that the robot could help her explain difficult concepts, \textit{``If I'm not sure if my explanation is getting through, I could say, `Let's ask the [robot],' and then we could listen to its explanation together.''} One participant (SC4) also suggested that the robot could provide information for blind visitors ahead of time, so it is easier for them to make requests or ask questions on the tour. This participant (SC4) mentioned that sighted participants can spontaneously ask a question by \textit{``look[ing] around''} in the environment; however, without visual information, the blind visitors can miss the chance to ask questions. As he suggested, \textit{``[The robot] can give a list, [stutter] basic information and the concept on the, on that permanent exhibition floor.'' (SC4).}

\paragraph{\textbf{Robot Design 2: group turn taking, human guide identification \& following}}
\revise{
Blind participants proposed various ways for the robot to support their group interactions, including finding and following the guide (P1, P3, P5), describing the group information (P4), and indicating turn-taking (P1). Three participants (P1, P3, P5) emphasized the need to identify available staff or experts to provide help, such as \textit{``getting explanations from a curator, or finding a staff member to get an explanation''} (P3). Similarly, P5 suggested, \textit{``Having a function to follow the tour guide would be very helpful.''} Another participant (P4) proposed that the robot could provide information about group situation, \textit{``If we stop and I wonder why, I can ask [the robot], `What's the situation now?' and it can tell me.''} The robot was also expected to answer questions related to group formation such as, \textit{``What's the current situation with the line [on the tour]?''} (P4). Furthermore, one participant (P1) suggested that the robot could suggest social turns in conversations with sighted people. As P1 explained, \textit{``Maybe when I want to ask a question in the middle of their talk, it’s hard to know how to ask a question nicely when there is a brief pause.''} This participant suggested using \textit{``vibration''} to indicate the turn for talking so the robot doesn't affect their hearing. 
}

SCs also proposed ways for the robot to facilitate blind visitors' interactions with the tour guide and other visitors. One participant (SC3) mentioned that the robot could encourage blind visitors to reach out to the human guide and ask questions, stating, \textit{``The [robot] knows what that SC is knowledgeable about, it could say so during the tour, `This person is an expert on this, so you should ask them about it.'''} Similar to one blind participant's suggestion (P1), SC4 suggested using vibrations to signal the conversation timing in case the visitors were too focused on one exhibit and lost track of the tour. As SC4 explained, \textit{``When the vibration happens, it means... like public conversation in the group starts or like the guide starts to give an explanation... Some people, when they are very much like concentrating on the exhibit, they forget about [stutter] getting some other information.''} (SC4).

\paragraph{\textbf{Robot Design 3: proximity and safety in navigation}} 

\revise{
Blind participants proposed the robot's navigational support for maintaining appropriate proximity within the group (P1, P4), enhancing safety and signaling dangerous situations (P2, P4, P5). First, the robot could help blind people maintain appropriate social distances from other people (P1) or form a circle in the group (P4). As P1 shared, \textit{``I can maintain a suitable distance from the people around me, without getting too close or too far away. I think I could follow them.''} Similarly, P4 mentioned, \textit{``When you stop and form a circle to talk, that kind of formation would probably be difficult.''} Second, participants suggested using the robot to promote independence (P5), ensure safety (P5) and give alerts in dangerous situations (P2, P4) during navigation. For example, P5 listed several tasks that blind people can perform independently with the robot's navigation support, \textit{i.e.}, \textit{``Not bumping into people, or being able to drink some coffee and take a break. Or going to the toilet and coming back. To be able to do these things without having to ask someone.''} In addition, participants (P2, P4) expressed concerns about violating social norms if the robot played loud sounds to give alerts during navigation. Instead, P4 suggested the use of light to signal alerts: \textit{`` I want it to be done subtly... If a lamp lights up, I think that would be enough. And if the museum staff or whoever sees that, if the lamp lights up, they'll know what it means.''}  Notably, one participant (P1) highlighted that the robot's navigational support can enhance their social interactions with nearby people by alleviating the movement challenges, stating, ``If I don't have to worry about movement or getting information, I'll have more mental space to talk with the people around me.''} (P1).

\revise{SCs shared several similar ideas with blind participants on the robot's navigational support, including} following the guide (SC1, SC4), alerting (SC1), describing the environment (SC1, SC4, SC5), and maintaining a safe distance from the exhibits (SC5). In addition, SCs suggested that the robot can provide directional information for blind visitors (SC5) and update the progress of the tour (SC1). As SC5 shared,\textit{ ``We can see various things in the [museum] while moving, but we can't explain them because we have to say things like, `We're going straight,' or `It's on your right.' So we might be skipping things [about the exhibit]. If the robot could handle that, we could divide the roles, and that might be better.''} Similarly, SC1 suggested the robot could update the progress of the navigation, \textit{``We have an oval bridge... We have come to one-third of the way... We have passed half of it... We have come to the end.''} (SC1). SC5 also described how the robot could help blind visitors to locate themselves by describing nearby people, \textit{``We could use the function that says, `Someone is so many meters away,' and have the [robot] say, `It's now so many meters away''' (SC5).}

\begin{table*}[t!]
\centering
\caption{The three major design goals were chosen based on the interviews with SCs and blind participants in the \textit{Interview Study}. Based on the design goals, we proposed robot features and user input methods to assist blind visitors' engagement in mixed-visual group tours. }
\label{tab-design}
\begin{tabular}{p{0.25\textwidth} p{0.36\textwidth} p{0.29\textwidth}}
\toprule
\textbf{Design Goal} & \textbf{Robot Feature} & \textbf{User Input} \\
\midrule
Environment Awareness  & 
\encircled[ff5e29]{1} Describe the nearby environment based on the semantic map and the live images captured by the robot's cameras & 
Press the LEFT button \\
\midrule
Group Interaction Awareness

& 
\encircled[ff5e29]{2} Describe the guide's position \newline 
\encircled[ff5e29]{3} Describe nearby people 
\newline
\encircled[ff5e29]{4} Notify the guide
& 
Press the RIGHT button 
\newline
Press and hold the RIGHT button
\newline
Press the DOWN button \\
\midrule
Independent Navigation
 & 
\encircled[ff5e29]{5} Always follow the guide and stop when the guide stops & 
User has no input \\
\bottomrule
\end{tabular}
\end{table*}

\section{System Design and Implementation}

\revise{Our interview study revealed rich design insights on how a robot could support group tours for blind people. Taking an initial step to explore the design space, we focused on three prevalent design goals and designed and implemented a robotic system to investigate people's interaction with the robot in real-world contexts. Specifically, the three design goals were blind visitors' environmental awareness, group interaction awareness, and independent navigation.}

\subsection{Apparatus}
The experimental system consists of an open-sourced robot platform\footnote{Cabot: \href{https://github.com/CMU-cabot/cabot}{https://github.com/CMU-cabot/cabot}} and two mobile phones for interaction. One mobile phone is attached to and connected to the robot on the user's end, hosting a \textit{User App}, and another is worn by the tour guide, hosting a \textit{Guide App}. The robot is used to support users' navigation, while the mobile phone provides auditory feedback to the user and facilitates communication with the guide. The user wears a bone-conduction headset to hear auditory feedback from the system through the \textit{User App}. The user also interacts with the robot through its handle, which has four mechanical buttons and three tactile protrusions that house a vibrator (Figure~\ref{fig:teaser}). 


\subsection{Interaction Design}

Table \ref{tab-design} describes the design of the guided tour. We generate ideas for the system design based on findings from the first study. These features serve as the first step in providing users with an experience of how the robot works and prompting them to generate further design ideas from the user study. 

\subsubsection{Environment Awareness} 
First, the system was designed to support the user's environmental awareness on the group tour by describing the surrounding environment. \encircled[ff5e29]{1} The user activates this feature by pressing the left button on the robot. Below is an example of the description of the surrounding environment:
\begin{quote}
   \textit{ This is an indoor exhibition space. On the left, an installation featuring branched wood serves as a space for reflection on the crisis of climate change and environmental issues. In front, there is a passage where people come and go, with wooden pillars and tables installed. On the right, there is a theater about stem cell research, such as iPS cells, which presents exhibitions that invite visitors to think about treatment options at the cellular level.}
\end{quote}

\subsubsection{Group Interaction Awareness}
Second, the system supports tour group interactions among the blind visitor, the guide, and sighted visitors by describing the position of the tour guide and nearby people, and notifying the guide. \encircled[ff5e29]{2} Users can request the location of the guide by pressing the right button on the robot handle. We differentiate the description based on whether the guide faces the user or not. The guide only faces the user when they are in front of the robot and look at the robot. The system will generate speeches to describe the guide's position, such as \textit{``The guide is facing you at 10 o'clock and is three meters away,''} or \textit{``The guide is three meters away, not facing you.''} \encircled[ff5e29]{3} Users trigger the nearby people description by pressing the right button on the robot's handle and holding it for one second. The description includes the number of people nearby and their distances to the robot, for example, \textit{``Three people are nearby, with one point five, two, and three meters away from the robot.''} \encircled[ff5e29]{4} The user can send a push notification to the tour guide by pressing the down button, and the guide would receive a verbal notification \textit{``Notifying the guide''} on the \textit{Guide App}. The guide can also send a message back to the user through the \textit{Guide App}, and the user will hear a verbal prompt, \textit{``The guide is coming''} on the \textit{User App}.
The user can press the up button to stop all robot speeches described above. 

\subsubsection{Independent Navigation} \leavevmode\encircled[ff5e29]{5} The user holds the handle on the robot to follow its movement. The robot always follows the tour guide and maintains a safe distance from other people and exhibits, so that the blind visitor can use the robot as a navigational aid to move independently. 




\subsection{System Implementation}
\begin{figure*}
    \centering
    \includegraphics[width=\linewidth]{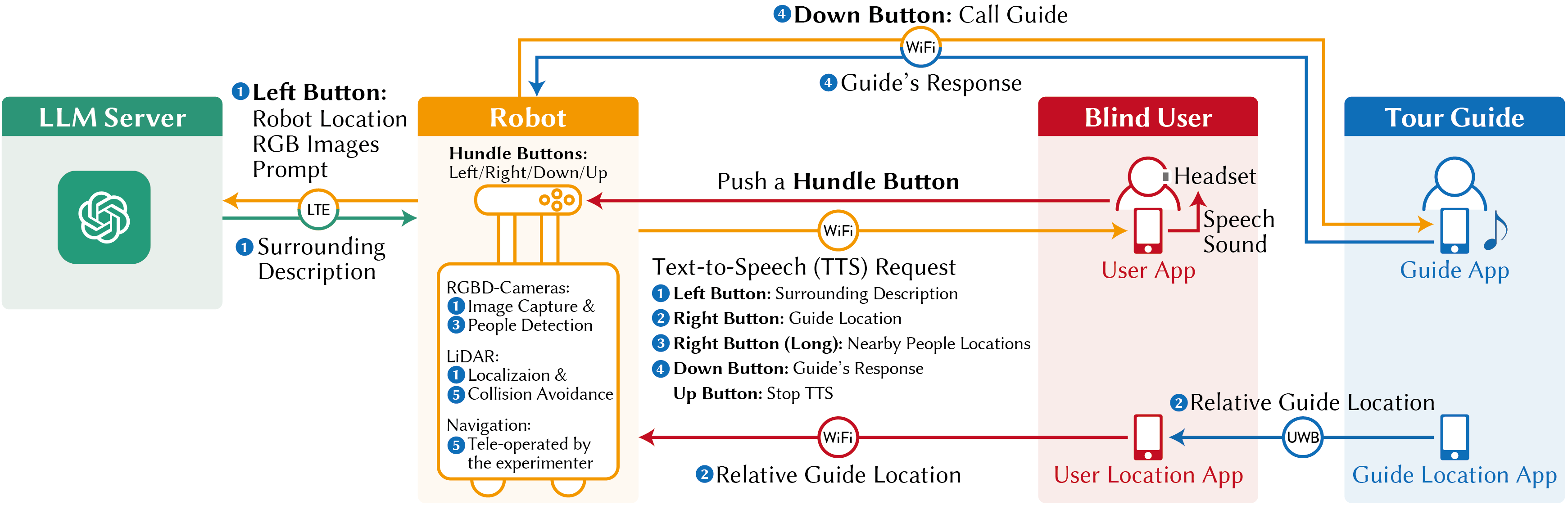}
    \caption{\textbf{System Overview.} The blind user presses the buttons on the robot handle and hears responses via a phone. The robot has five main features to support the blind visitor's group tours: 1) The robot uses its cameras to send images and its current position to the LLM server, and receives a surrounding description; 2) The robot's phone receives UWB broadcasting signals from the phone carried by the guide, and the robot determines the location of the guide; 3) The robot uses its cameras and a local detection model~\cite{mmdetection} to detect nearby people; 4) The robot triggers a notification on the phone carried by the guide; and 5) The robot navigates via tele-operation while using its LiDAR for collision avoidance.}
    \Description{A system diagram showing how each component in the system communicates with each other. 
Four main boxes from left to right are: LLM Server, Robot, Blind User and Tour Guide 

1. LLM Server connects via LTE to the Robot, providing surrounding description with inputs of Robot Location, RGB Images and Prompt.
2. The Robot has handle buttons (left, right, down, up). Each button triggers a text-to-speech (TTS) request:
Left: surrounding description
Right: guide location
Right (long): nearby people locations
Down: guide’s response
Up: stop TTS
Inside the robot figure, there are the following texts:
RGBD-Cameras: 1. Image Capture & 3. People Description 
LiDAR: 1. Localization & 3. Collision Avoidance 
Navigation:
5. Tele-operated by the experimenter 

The Blind User interacts with the robot through a user app on their phone and hears audio via a headset. 
Under user, there are the User app and User location app.
Under tour guide, there are the Guide app and Guide location app. 
Guide Location App  sends relative guide location data to the User Location App via UWB.
Communication pathways are illustrated:
LTE connects the robot and LLM server.
WiFi connects between the robot and user app.
WiFi connects  between the robot and the guide app.
UWB connects user location app and guide location app.
}
    \label{fig:system-design}
\end{figure*}

The robot provides a WiFi network for WebSocket communication between the robot and the mobile phone apps. In the following, we introduce the implementation for each component of the system and the communication between them (Figure~\ref{fig:system-design}).  

\paragraph{Surrounding Environment Description} \encircled[1D70C1]{1} The robot generates a surrounding description utilizing a vision language model (GPT4o) \footnote{GPT4o: https://openai.com/index/hello-gpt-4o/}. The description of the surrounding environment was based on nearby exhibits and three images captured by the three cameras (left, front, and right) on the robot. The model combines data from the three RGB images along with descriptions of nearby exhibits based on the robot's location and orientation. The experimenter collected the location and angles of exhibits on the map and prepared descriptions based on the human-guided tour scripts and exhibition information on the museum website. Once the user triggers the generation of the surrounding environment, the robot first searches for exhibits within a 20-meter radius. If any exhibit is found, the system references the robot's current angle to identify the nearest exhibit in the forward, right, and left directions. Their description text, together with the relative positions regarding the robot's heading, is fed into the prompt of the vision language model. In addition, three real-time images captured by the robot's cameras are added to the prompt, specifying whether the image was on the left, right, or in the front. After generating the description on the robot's server, the message will be sent to the \textit{User App}. The prompt information is reported in the Supplementary Materials.

\paragraph{Guide Location and Orientation} \encircled[1D70C1]{2} We developed the \textit{User Location App} and the \textit{Guide Location App} to obtain the relative locations of the guide from the robot and the user. We used two additional mobile phones to easily integrate the ultra-wideband (UWB) functionality; however, these functionalities should be merged into the \textit{User App} and \textit{Guide App} used for interactions and communications between the user and the guide.

\paragraph{Nearby People Location} \encircled[1D70C1]{3} The system detects people in proximity through three RGB-D cameras mounted on the front, left, and right sides of the robot. The MMDetection~\cite{mmdetection} system is used to detect and segment people in images and estimate the distance from the camera. After generating the nearby people description, the message is sent from the robot's server to the \textit{User App}.

\paragraph{Communication with the Guide}
\encircled[1D70C1]{4} The down button on the mechanical handle is used to send the push notification to the guide, and the \textit{User App} receives the response from the guide. The \textit{Guide App} sends a message and receives the notification from the user. All the communication is over WebSockets, which is hosted by the robot's server.

\paragraph{Robot Navigation}\encircled[1D70C1]{5} 
Due to the prototype nature of the system and the challenges of ensuring robust robot following in a crowded environment, the robot's navigation was tele-operated by the experimenter to ensure that it always followed the guide. Although the robot was tele-operated, it attempted to avoid collisions with other participants and obstacles through a safety control system based on the LiDAR measurements and people detection results.

\begin{table*}
\caption{
\textbf{Blind Participant Demographics.} The mean age was 51.1, with a SD of 14.8. All participant used a cane as their daily navigation aid. None of the participants has participated in tours offered by this museum.
}
\label{tab:demographics_blind_main}
\small
\begin{tabular}{ccccccc}
\toprule
\multirow{2}{*}{\textbf{ID}} & \multirow{2}{*}{\textbf{Gender}} & \multirow{2}{*}{\textbf{Age}} & \textbf{Impairment} & \textbf{Tours} & \multirow{2}{*}{\textbf{This Museum Visit}} & \multirow{2}{*}{\textbf{Used Similar Robots}} \\
                             &                                  &                               & \textbf{Duration}            & \textbf{Experience}     &                                             &                                              \\ \hline
P01                          & M                                & 21                            & 19 years            & Y              & 5+ times                                    & 3 times                                      \\
P02                          & M                                & 65                            & 25 years            & Y              & 1 time                                      & 1 time                                       \\
P03                          & M                                & 60                            & 30 years            & Y              & 2 times                                     & 3+ times                                     \\
P04                          & M                                & 33                            & 33 years            & Y              & 10+ times                                   & 5+ times                                     \\
P05                          & M                                & 57                            & 15 years            & N              & 1 time                                      & 0 times                                      \\
P06                          & F                                & 64                            & 16 years            & N              & 10 times                                    & 3 times                                      \\
P07                          & F                                & 52                            & 20 years            & Y              & 10+ times                                   & 5 or 6 times                                 \\
P08                          & F                                & 56                            & 30 years            & Y              & 10+ times                                   & 4+ times                                     \\ \hline
\bottomrule 
\end{tabular}
\end{table*}

\section{Field Study: Robot-Assisted Group Tours in Public}
We conducted the second user study where blind visitors joined guided group tours with sighted visitors. The tour guide was a museum staff member and conducted tours based on scripts prepared for the tour routes. In each study session, one blind participant and two sighted participants joined the tour with the tour guide. A male museum staff member served as the tour guide. He was 63 years old and had 20 years of experience managing the museum, along with several years of experience conducting tours for sighted visitors. The tour was designed with advice from SCs who had experience leading tours for blind visitors, to ensure that minimum accessibility requirements were met.

\subsection{Participants}
We recruited eight blind participants and eight sighted participants for the study. Blind participants were recruited using the same mailing list as the interview study, and none of them had participated in the interview study. The detailed demographic information is shown in Table~\ref{tab:demographics_blind_main}. The sighted participants were recruited from a local university, and all were unfamiliar with the exhibition site. The group consisted of seven males and one female, with a mean age of 23.63 years (SD = 1.87). None of the sighted participants has participated in tours with blind people before.

\subsection{Study Setup}
\paragraph{Tour Activities}
The guided group tours were conducted in a science museum with a theme of \textit{Earth} and the universe. Two sets of tours were prepared for the study and each tour went through three \revise{different} exhibition areas. \revise{Each tour consisted of one blind visitor and two sighted visitors. Sighted participants joined both sets of tours. Each blind participant joined one of the two sets. As sighted visitors did not directly use the robot, participating in two sets of tours increased their exposure to the robot, and they experienced blind participants with different interaction styles.}   


\subsection{Study Procedures}
The field study consisted of pre-tour sessions, including study briefing and robot training and practice, group tours with the robot, and interview sessions with both blind participants and sighted participants. Below, we first report the study procedures for blind participants and then for sighted participants. All interview sessions were audio recorded and the tours were video recorded. 

\subsubsection{Session with Blind Participants}
\paragraph{Pre-tour Session}
The experimenters met the blind participant at the science museum or the nearby bus/train stations. The pre-tour sessions were hosted in a conference room in the museum. We first introduced the study process and went through the consent form with the blind participant. After participants signed the consent form, we interviewed them about their past experience in group tours. The pre-tour session lasted for around 15 minutes.

\paragraph{Robot Introduction and Training}
After the pre-study interview, we conducted the robot introduction and training session in the museum exhibition area. One experimenter first guided the blind participant to touch each hardware component of the robot while providing explanations. Next, the experimenter introduced the handle buttons on the robot and the participant triggered the description of the surrounding environment, nearby people, and the tour guide's position. The participant also practiced stopping the robot's speech during its description. After the robot introduction, the participant practiced the robot use by joining a mini tour with the guide. During the mini tour, the participant used the robot to navigate and follow the guide. They were also asked to press the four buttons on the robot handle to hear the robot's speeches on the mini tour. The blind participant practiced the use of the robot until they felt comfortable using the robot. The robot introduction and training session lasted for 15--30 minutes.

\paragraph{Tour with the Robot}
After the training session, the two sighted participants joined the tour group with the blind participant. The guide first welcomed the visitors to the tour. Then, during the tours, the tour guide explained the exhibitions, asked quiz questions to the group, and guided the blind participant to experience the touchable elements on the tour. The tour guide used blind visitor-friendly language to introduce the exhibits. For example, the tour guide described the size, shape and color of the exhibits, whether there was a glass in front of the exhibit, and whether the exhibits were touchable. The tour lasted for 30--40 minutes. 

\paragraph{Post-tour Session}
After the tour, we conducted post-tour sessions with the blind participant. Interview questions included their overall tour experience, feedback for the robot features, and whether the robot supported the three needs we identified from the interview study (Table~\ref{tab-design}). Participants also provided ratings for SUS~\cite{brooke1996sus} and RoSAS~\cite{carpinella2017rosas} scales, and provided demographic information. The post-tour session lasted for 30--40 minutes. 

\subsubsection{Session with Sighted Visitors}
Before the tour, we briefly introduced the study to sighted visitors and informed them that there would be a blind visitor using a robot in the guided tour. They were asked to engage in the tour in the same way as they did in their past tours. After the tour, we conducted a group interview with the two sighted participants to collect their feedback on participating in the tour with the blind visitor and the robot. The post-tour interviews with sighted participants lasted for 10--15 minutes. 

\subsubsection{Interview with Tour Guide}
After all eight field study sessions, we conducted one interview session with the tour guide and asked questions about feedback for the tour and the robot. The interview session lasted for 20 minutes.

\begin{figure*}[t!]
    \centering
    \includegraphics[width=\textwidth]{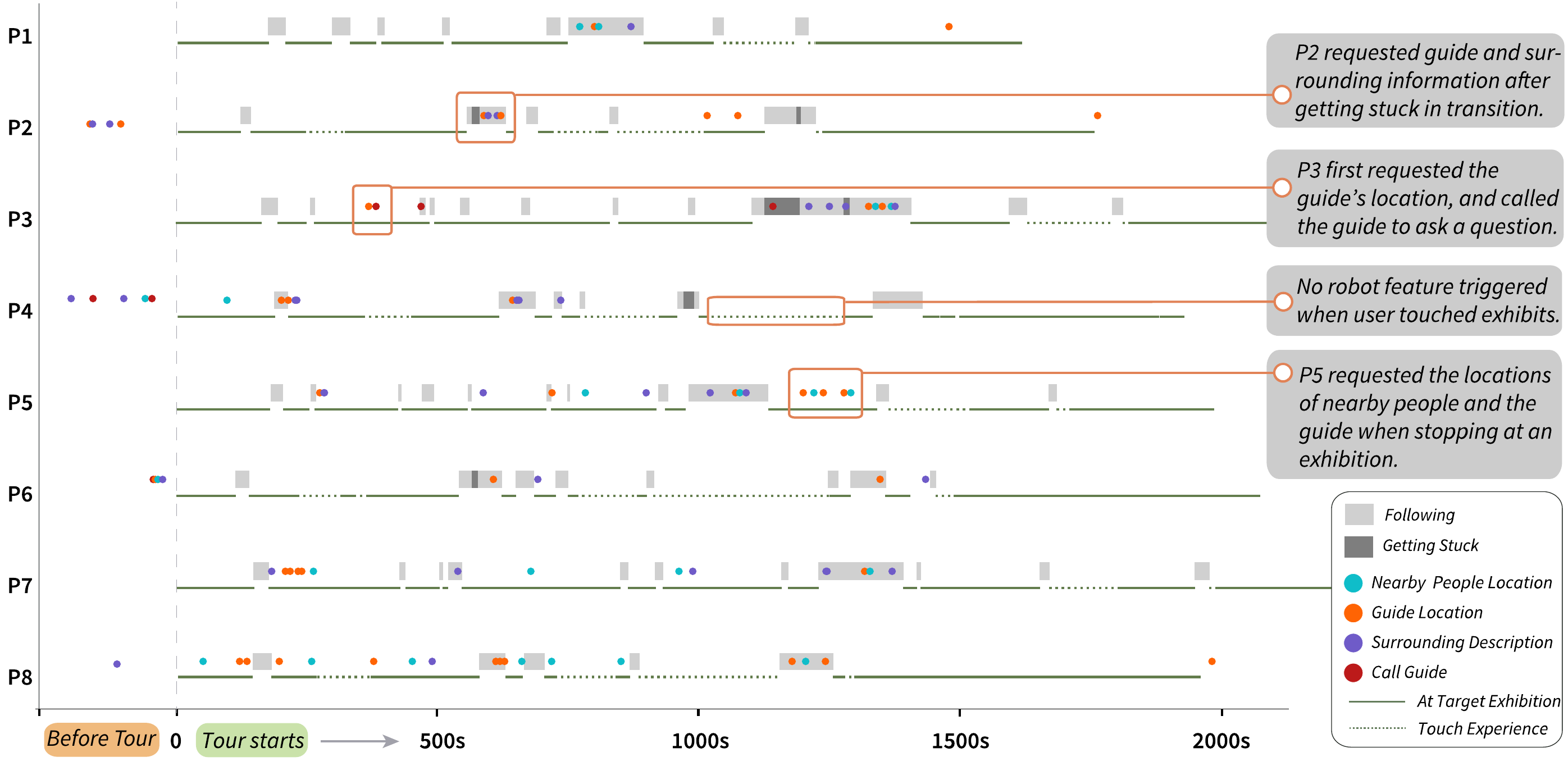}
    \caption{\revise{\textbf{Tour timelines with 1) tour stages, 2) robot navigational behaviors, and 3) features triggered by users.} On the tour, the group visited a fixed set of exhibits (\textbf{solid line}) guided by the tour guide and had touch experiences (\textbf{dotted line}). The space in between the lines represents the transition from one exhibit to the next. The user used the robot to follow the robot (\textbf{gray block}) and can get stuck sometimes due to crowds or obstacles (\textbf{dark gray block}). The timeline also illustrated the four features triggered by the user before, during, and after the tour.}}
    \Description{Timeline illustrating the tour group states, features triggered by the user and robot navigational behaviors.}
    \label{fig:video-analysis}
\end{figure*}

\subsubsection{Data Collection and Analysis}
We collected and analyzed three types of data, including interviews with blind and sighted participants, system logs that captured interactions between the user and the robot and videos of the tours. 

\paragraph{Interview Data}
We conducted reflective thematic analysis \cite{braun2019reflecting, byrne2022worked} of the interview data following the same procedure as the interview study analysis as reported in \S\ref{formative-analysis}. \revise{Analysis of the blind participants' interviews, the sighted participants' interviews, and the tour guide interview were conducted following the same protocol. After all datasets were coded, the lead author first created initial themes from the codes and then discussed them with the research team to finalize the final set of themes.} 



\paragraph{System Logs}
System logs captured all instances when the user pressed buttons on the handle and the robot's verbal responses. \revise{The system logs also included the timestamp of each instance.} To understand how participants used the robot, we calculated users' frequencies of triggering each robot feature. 

\paragraph{Video Analysis}
\leavevmode\revise{To understand when and how blind users used the robot on the tour, we analyzed tour videos to identify tour stages and robot behaviors in each tour stage. As the tour was designed with a fixed set of exhibitions and touch experiences on the route, we took a deductive thematic analysis approach \cite{braun2006using} and coded the tour with five stages: \textit{Before the Tour,\footnote{After the training, blind participants would wait for the tour to start at the entrance}} \textit{At the Exhibit,} \textit{Touch Exhibit,} \textit{Transition between Exhibits,} and \textit{After the Tour.} We coded the robot's navigation with two states, \textit{i.e.}, \textit{Following} and \textit{Getting Stuck}, to understand users' navigational experiences. In addition, we mapped robot features triggered by the user from system logs into the tour video's timeline and identified group activities surrounding the feature use. The lead author performed the video coding and analysis using an open-sourced tool ELAN\footnote{ELAN: https://archive.mpi.nl/tla/elan} and the research team reviewed the results.} 


\section{Findings}
\subsection{Overall experience and system usage}

The majority of our blind participants reported positive experiences in the robot-assisted group tours. SUS scores ($M=90.6, SD=12.9$) indicated excellent system usability. Seven out of eight participants (P2--P8) provided positive feedback, described the experience as \textit{``surprising''} (P3), \textit{``enjoyable''} (P4, P6) and \textit{``fun''} (P3, P8). However, one participant (P1) reported a negative experience and mentioned the difficulty in \textit{``keep[ing] up with the understanding''} with sighted visitors and viewed the tour \textit{``the same as sitting in a seat and listening to a talk''} (P1).  RoSAS scores also indicated positive social attributes, with high ratings on positive attributes (\textit{e.g.}, interactive, reliable, responsive, competent, and knowledgeable) and low ratings on negative attributes (\textit{e.g.}, scary, strange, awful, awkward, dangerous, and aggressive).

\begin{figure*}
    \centering
    \includegraphics[width=\linewidth]{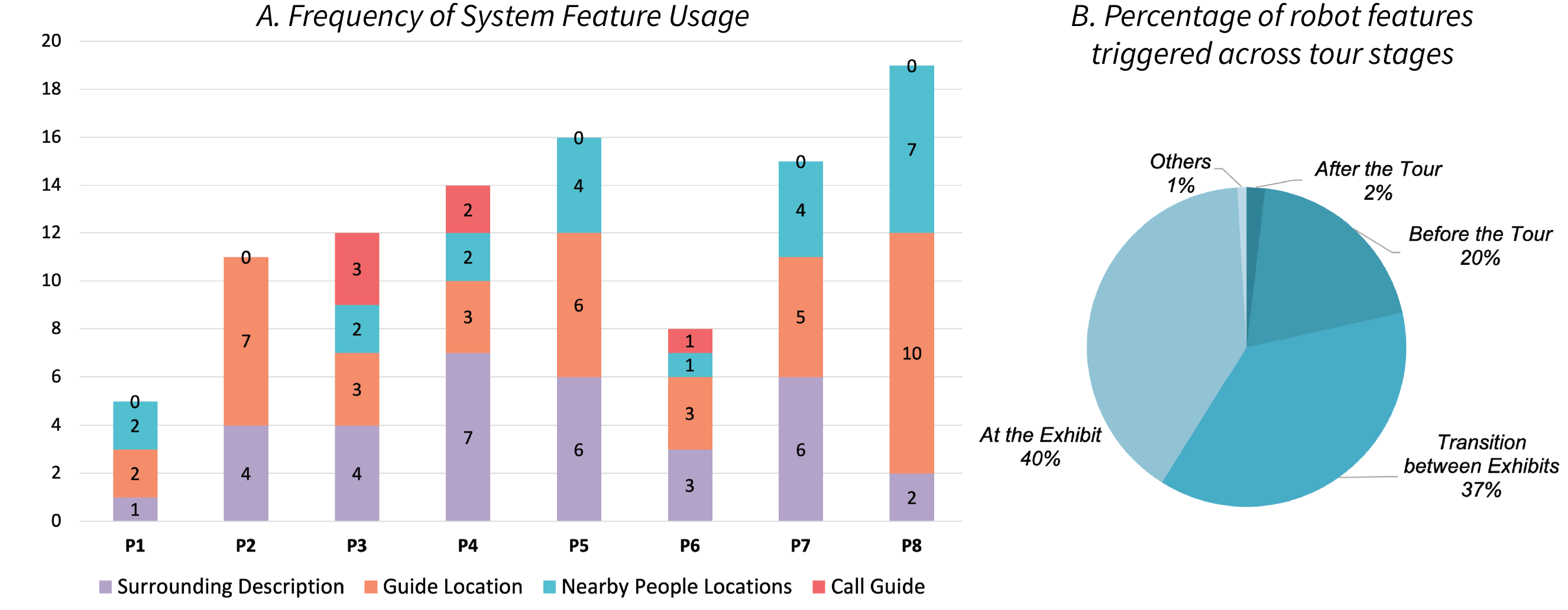}
    \caption{\textbf{Left:} How many times each feature was triggered during the tours experienced by participants? The mean and SD of each feature usage are: \textit{Surrounding Description} (Mean 4.13, SD 2.10), \textit{Guide Location} (Mean 4.88, SD 2.70), \textit{Nearby People Locations} (Mean 2.75, SD 2.19), \textit{Call Guide} (Mean 0.75, SD 1.16). Additionally, the stop feature, which mutes the robot, was triggered with a mean of 7.62 and a SD of 6.20. As reported by the participants, because there is a slight delay between triggering this feature and muting the robot, they would often trigger it repeatedly, resulting in a high count. \revise{\textbf{Right:} Percentage of tour stages when participants used the robot's features. }}
    \Description{The figure describes a bar chart and a pie chart. On the left, the title reads “A. Frequency of System Feature Usage”. The units of the numbers are counts. The horizontal axis is the participant IDs from P1 to P8. The vertical axis is the quantity scale at 0, 5, 10, 15, and 20. The legends are pink - surrounding description, yellow - guide location, purple - nearby people locations, and green - call guide. From P1 to P8, the counts for the pink-surrounding description are 3, 4, 4, 7, 6, 3, 6, 4. The counts for the yellow-guide location are 2, 8, 3, 3, 6, 3, 5, 10. The counts for the purple - nearby people locations are 2, 0, 2, 2, 4, 1, 4, 7. The counts for the green - call guide are 0, 0, 3, 2, 0, 1, 0, 0.  On the right, the title reads "B. Percentage of robot features triggered across four stages". Transition between exhibits is 37\%, before the tour is 20\%, after the tour is 2\%, others is 1\%, at the exhibition is 40\%.}
    \label{fig:usage log}
\end{figure*}

\subsubsection{System Usage Pattern}
\revise{
Our video analysis demonstrated system use patterns by blind participants. As illustrated in Figure~\ref{fig:video-analysis} and Figure~\ref{fig:usage log}, robot features were triggered predominantly with similar frequencies when the tour group stopped at exhibitions (\textbf{39\%}) and when transitioning between exhibitions (\textbf{37\%}). Participants revealed different preferences in the timing of using robot features. Three participants (P1--3) primarily triggered robot features during transitions, while three participants (P5, P7, P8) frequently used them across the whole tour. Also we observed that three participants (P4, P6, P7) tended to request the group's location and surrounding information when arriving at a new exhibition. Participants (P2, P3, P6) also requested group locations and surrounding information when the robot was stuck, and one of them (P3) called the guide after the incident occurred. Regarding social interactions, while most participants asked the guide questions directly or used gestures to get the guide's attention, only one participant (P3) used the robot to get attention from the guide. This participant first requested the guide's location and then used \textit{Call Guide} to get the guide's attention and ask questions.}  

\revise{Participants also used the robot features during stages where there were uncertainties about the situation, such as right before (\textbf{20\%}) or after the tour (\textbf{2\%}). Four participants (P2, P4, P6, P8) requested the group and surrounding environmental information while waiting for the tour to start. One of them (P4) tried to call the tour guide twice before the tour. Two participants (P2, P8) triggered the ``Guide Location'' feature right after the tour ended. One participant (P6) also requested surrounding information in an exceptional case (\textbf{1\%}) when the guide walked away to pick up gadgets.}

\revise{We also observed that the robot got stuck from time to time in several tours (P2--4, P6), as illustrated in Figure~\ref{fig:video-analysis}. These incidents were primarily caused by crowds in the museum, especially for afternoon tours during peak hours. The robot freezing issue can also occur due to the user's positioning, such that the robot may mistakenly detect the user's feet or the accompanying white cane as obstacles. 
}

\subsubsection{System Usage Frequency}
 As illustrated in Figure~\ref{fig:usage log}, the ``Guide Location'' feature was triggered the most among the main features, with all participants using this feature. The ``Surrounding Description'' feature was also triggered with similar quantities. Participants in general favor these two features because, as discussed in \S~\ref{sec:robot-assisted-group-interaction}, these features offer vital information for the participants to participate in group interactions. Except for P8, all participants triggered comparatively fewer ``Nearby People Locations'' feature. As discussed in \S~\ref{sec:interaction-with-other-visitors}, participants wish to see further enhancements of this feature for it to be useful. The \textit{Call Guide} feature was triggered the least. However, as discussed in \S~\ref{sec:maintain-connection-with-guide}, participants reported not needing to use this due to the small size of the tour groups and considering this feature valuable in large groups or during emergencies. Finally, the varying distributions of features triggers across participants reflect diverse demands for the robot's assistive features.

\subsubsection{Comparing Robot with Human Companion}

Participants (P1, P4, P6, P8) compared the companionship from the robot with that from a human. For participants who favored the robot companion over a human companion, the robot enabled them to join the tour independently (P6), allowed for more flexible schedules (P4), and control over the experience (P8). For example, P4 reported the difficulty in coordinating with a human companion, \textit{``When it comes to people, you have to arrange a meeting, coordinate schedules, and participate like that. However, when participating with a robot, as long as it fits your schedule, you can just casually go whenever you have time, and I think that's a good thing.''} Another participant (P8) shared that she has more control over the experience (\textit{e.g.}, when to talk) with the robot, saying that \textit{``If you are with someone, especially family, they might start talking to you, and you can't listen to the explanations properly.''} On the other hand, one participant (P1) preferred to have a human guide than the robot because he could ask questions more easily, commenting, \textit{``While the explanation was going on, I could casually ask, `What kind of exhibit is this now?' or `Can I touch this?'''} (P1).  




\subsection{Robot-Assisted Group Following}

\subsubsection{Sense of Safety in Navigation}
Participants (P2--P7) reported an increased sense of safety with the robot's assistance. With the robot's navigation support, participants mentioned that they felt \textit{``safe''} (P4) and walked with \textit{``peace of mind''} (P2, P3, P5). Knowing the distances from nearby people from the robot's description (P5) and being able to notify the guide (P7) also enhanced the sense of safety. As P7 commented, \textit{``Even if you got lost, you could participate with confidence.''} (P7).
In addition, participants proposed robot features to further enhance navigation safety by describing children nearby and supporting awareness of the side of the body opposite the robot’s handle. 
One participant suggested that the robot should explicitly describe the presence of nearby children, noting that children can cause safety risks due to their unpredictable movements: \textit{``When children are around, they tend to act a bit unpredictably, like running around and such.''} (P6). 
Another participant (P2) expressed concern about the safety of the side of his body opposite the robot, as he did not know the distance to surrounding objects or people and had to adjust his posture while walking. As he explained, \textit{``I kept wondering how much space there was on the left... I ended up kind of narrowing my body, or turning a little sideways, not facing straight ahead.''} (P2). 


\subsubsection{User Control and Robot Feedback in Guided Navigation}
In our current design, the robot always follows the tour guide without any user input. However, participants suggested ways for the user to control the robot (P2, P3) and receive navigation-related feedback from the robot (P1, P4, P6). One participant expressed the desire to control the robot to move closer to the guide when he ``felt like there was a bit of distance'' (P3). Two participants emphasized the need to adjust the robot's speed \textit{``for older adults''} (P2) and when \textit{``falling behind''} (P3). Participants also suggested that the robot could indicate its action before moving, as they experienced sudden robot movements. They proposed using \textit{``vibrations''} to signal the movement changes (P1), providing \textit{``brief guidance''} (P4), and using verbal commands such as \textit{``this way''} (P6) before moving. 



\subsection{Robot-Assisted Group Interaction}
\label{sec:robot-assisted-group-interaction}

\subsubsection{Engagement with the Tour Guide} 
On the group tour, the guide's explanation was the primary way for blind visitors to receive new information about the exhibition. Below, we present how the robot can supplement the guide's explanation as well as pose risks of distracting visitors' interaction with the guide. 

\paragraph{Supplementing the Guide's Explanation}
Participants(P3, P4, P6) described how the robot provided different information from the human guide. For example, P6 reported that the human guide talked about \textit{``the theme of the guided tour''} while the robot focused on the \textit{``descriptions of the surrounding situation''} (P6). Another participant (P3) shared how the robot gave a preview of the exhibit before the guide's explanation, \textit{``It provides information that is not covered in the guide’s explanation, or lets me confirm, `Ah, this is how they’re describing what was just explained.'''} (P3). P4 appreciated that she can obtain the information \textit{``by combining the, the person and the robot''} and can choose to listen to the robot or the human guide by pressing the stop button. 
In addition, participants (P6, P7) proposed that the robot could augment the human guide's explanation. For example, the robot could give a high-level summary of nearby exhibits so the user could ask the human guide for more detailed information (P7). The robot could give explanations again if the visitor missed any information (P6). As P6 described, \textit{``I think it would be nice if there were a way to say, `I just want to hear this part again, this specific bit.'''} (P6). 


\paragraph{Increased Attention to the Guide}
The robot's support could lower blind visitors' mental load during navigation so they could be more focused on interacting with the guide (P2). For example, one participant (P2) compared his experience using the robot with using a white cane, saying, \textit{``When using a white cane, I'm constantly exploring and probing on my own, but without it, just being able to follow along is very comfortable... It means you can focus on listening to the guide's talk.''} (P2). 

\paragraph{Distraction from the Robot}
Nevertheless, interactions with the robot could impair the tour experience as participants (P3, P4, P7) reported overlaps between the interaction with the robot and that with the human guide. For example, P4 described how controlling the robot distracted him from listening to the guide, \textit{``I was trying to stop the explanation abruptly, but while I was doing that, the guide's talk went a bit further, and I ended up missing a part of it.''} As a result, several participants (P1, P6, P8) chose to prioritize the interaction with the human guide and minimize the use of robot features to avoid distractions. As P1 mentioned, \textit{``Instead of wanting to know the entire surroundings, I was more interested in listening to the guide's explanations.'' }(P1).

\subsubsection{Maintaining Connections with the Guide}
\label{sec:maintain-connection-with-guide}

\paragraph{Needed in Big Groups and for Emergencies}
The robot had a \textit{Call Guide} feature that allowed users to send a push notification to the guide if they had questions or needed help. Although we observed low usage of this feature in the field study, participants mentioned that this feature was necessary and would have been more helpful if the group were larger (P1, P5, P6--P8). In addition, participants (P5, P6) suggested that they would like to use the feature when an emergency occurs, \textit{e.g.}, for requesting bathroom use or resting. As P6 explained, \textit{``If I want to say something like, 'My foot hurts a bit, excuse me, guide, could you please stop for a moment,' but if I'm far away, I can't say it.''} (P6). 

\paragraph{Alternative Communication with the Guide}
Although our current system only supported push notifications to connect with the guide, participants proposed alternative ways for communicating with the guide through the robot. First, the robot could send a subtle reminder to the human guide without drawing much attention. For example, the robot could ask the guide to speak louder (P3), repeat what they just said (P3, P7), slow down when the blind visitor fell behind (P3), and \textit{``stop for a moment''} (P5). For example, P3 explained the necessity of these subtle reminders, saying, \textit{``It’s not something worth calling the guide over just for that, but it’s something I’d like to communicate... Things like `speak louder' or `please repeat that.'''} With these reminders, participants described the robot as a proxy that \textit{``speak on your behalf''} (P5) and \textit{``speak for me''} (P7). Furthermore, one participant (P8) proposed talking to the guide directly through the robot with \textit{``the function to call.''} As P8 explained, \textit{``I can imagine situations where I can't ask questions, so I think it would be nice to have a call function that allows me to ask questions.''} (P8).


\subsubsection{Interaction with Other Visitors}
\label{sec:interaction-with-other-visitors}
\paragraph{Interaction Preferences}

Participants did not actively interact with other visitors and primarily focused on the guide's explanation during the tour. Limited social interactions in the group included answering quiz questions from the guide and back channeling, \textit{e.g.}, nodding to respond to others and using \textit{``ah''} or \textit{``yeah''} (P7). For example, P2 described the quiz interactions, saying, \textit{``When we have a quiz or something and we listen to everyone's opinions.''} Participants reported challenges in talking to people they did not know, \textit{e.g.}, \textit{``I just couldn't bring myself to speak up.''} (P1) and \textit{``I am not very good at talking to people for the first time.''} (P8). 

While most participants had limited active social interactions within the tour group, two participants (P4, P5) reported the importance of sharing thoughts with other people and hearing their opinions on the tour. As P4 mentioned, \textit{``Since we're in the same space, I feel like there might be some common topics to talk about, so I'd like to try having a conversation if something comes to mind.''} Similarly, P5 emphasized the need for communication with other visitors: \textit{``Since we're together, I think it's important to talk.''} (P5).

\paragraph{Robot Description of Nearby People}


Although not actively interacting with nearby people on the tour, participants frequently triggered the \textit{Nearby People Location} feature on the robot (Figure~\ref{fig:video-analysis}). The feature was used mostly when participants were on the move from one exhibit to another exhibit (P1, P3, P5, P7, P8) or when they stopped in front of the exhibit (P4, P5, P7, P8). One participant (P8) commented that knowing the distance helped her better communicate with other people nearby: \textit{``By knowing the distance, you can judge whether your words will reach them if you speak to them now.''} (P8). Another participant (P5) used this information to infer the group dynamics, stating that, \textit{``If it said there are three people at some point, but earlier there were four or five, I could guess that maybe two of them went off in another direction or are having a separate conversation.''}(P5). 

In addition to distances, participants (P2, P3, P5, P7, P8) proposed additional descriptions of the nearby people, such as their orientations, genders and physical appearances. For example, one participant (P3) suggested describing whether nearby people are on their left or right, stating, \textit{``Since I couldn’t tell the left–right positioning, that was a bit of an uncertainty.''} In addition, participants requested information about nearby people's gender (P2, P7, P8), facial expressions (P5) and identity (P3, P5). In particular, P3 mentioned these detailed descriptions of nearby people could help them find their companions, \textit{``If I were with family or friends, and I happened to be relying on the robot, then I think it would be good if I could know the situation of the people I was with.''} (P3).

\subsubsection{Heightened Self-consciousness}
Participants (P1, P8) pointed out the differences between blind and sighted people and reported heightened self-consciousness in joining mixed-visual group tours. For example, one participant (P1), who had a negative experience, mentioned the feeling of \textit{``being different''} and the difficulty in catching up with the group in recognizing and understanding the exhibition. As P1 commented, \textit{``I was the only one who somehow couldn't quite keep up with understanding.''} Similarly, another participant  (P8) shared the self-consciousness of being treated specially, \textit{``I felt like they were being quite considerate towards me, so I wondered if others might think that I was being given special treatment.''} (P8).

\subsection{Feedback from Sighted Stakeholders}
\label{sec:stakeholder-feedback}
\paragraph{\revise{Sighted} Participants' Feedback}

All sighted participants (SP1--SP8) reported a positive experience on the tour and described it as \textit{``interesting''} (SP3, SP4) and \textit{``enjoyable''} (SP6). Three participants (SP4, SP7, SP8) highlighted the touch experience for the exhibition on the tour. For example, SP8 highlighted the benefits of the touch experience compared to the visual presentation, stating \textit{``When you touch it... Like the CO2 emissions, they're also written in numbers, but I thought it was more intuitive and easier to understand if they were represented with balls.''} On the other hand, sighted participants reported challenges they faced on the tour with the robot, particularly that the delay in the robot's movement led to a slower pace of the tour (SP3--SP5, SP7, SP8). For example, one sighted participant (SP8) expressed the confusion of positioning themselves with respect to the robot on the tour, saying, \textit{``I was a bit confused at first about whether it was okay to overtake or if I should walk behind.''} Sighted participants (SP3, SP4, SP7) also reported that they tended to be more careful and tried to avoid the robot when walking. As SP7 commented, \textit{``It might be better to avoid getting in the way of the [robot] or crossing its path too much. So, I was a bit careful of the route I took.''}  





\paragraph{Tour Guide's Feedback} 
The tour guide emphasized the need for the \textit{Call Guide} feature, in particular if the group were bigger. As he mentioned, \textit{``When my voice couldn't be heard, I think someone pressed a button, so if it's used that way, I can go closer to them.''} The guide also described how he accommodated the robot on the tour by monitoring the robot and adjusting the speed. As he reported, \textit{``If it fell behind, I would adjust my speed.''} The guide also suggested additional features to monitor the robot and shared how it can be used to ensure safety in a larger space, \textit{``First of all, we can know where it is, right? And for the tour, maybe in a factory, not [the museum], but like a factory tour with a large group in a wide space, like I mentioned, if they get lost, it would be great to be able to press a button and find out where they are.''} (SP7).

\section{Discussion}


Participating in group activities with people with mixed vision abilities is challenging for blind people. Using a guided group tour as a scenario, we investigate how an assistive robot can support blind people to follow the group and interact with people on the tour. One key challenge for blind people's participation in tours is the anxiety of navigation and interactions in groups. Our field study demonstrated that group tours can be enjoyed by blind visitors and sighted visitors together. The robot could increase blind visitors' sense of safety in participating in the tour independently and support their engagement with the guide to obtain new knowledge. The touching and tactile experiences designed for blind visitors could also create new experiences for sighted people. Our participants reported an overall positive tour experience with the robot.  

\subsection{Design Implication}
\subsubsection{Needs and Challenges for Blind People Tours}
In answering \textit{\textbf{RQ1:} What are blind people's needs and challenges in group participation?} We interviewed blind people and museum experts from a science museum. Our findings revealed that group tours pose challenges for blind visitors to navigate with the group, keep up with the tour pace and maintain connections with the guide. As shared by our museum expert participants (SCs), accessible tour design for blind people should include rich touch experiences, provide visual information description, and manage the group flow in the crowds to ensure safety. Existing group tours for blind people in the science museum require longer durations and more museum staff support, as a result, these tours occur at a lower frequency compared with regular tours. Our SC participants proposed various ways that assistive robots can support blind people and share the workload for the tour guide, with respect to navigational, informational and communicational support. 

One suggested robot feature is to inform the blind visitors about the path that the group will be navigating when going to the next exhibit. It points to an important need for inferring group intentions and next-step actions. Future work can investigate how to obtain and convey this information and how it might affect the situation awareness and group interactions for blind people. For example, in the museum tour scenario, the robot need to combine information of the guide's plan, group dynamics and the surrounding environment to predict the next navigation destinations. In addition to the real-time sensory input and pre-stored environmental information, the robot may interact with the guide and other people in the group to obtain planned activities. 


\subsubsection{Lower Barriers for Accessible Tour Experience}
In answering \textit{\textbf{RQ2:} How can we support these needs through the design of a mobile robotic assistant?} We designed and implemented assistive features on a mobile robot that supports environmental awareness, group interaction awareness, and independent navigation for blind visitors on group tours. Findings from a field study highlighted how the robot assisted blind participants' group following and group interaction on a tour. The robot supported the sense of safety for blind people through supporting navigation, providing environmental information, and keeping in connection with the guide. This benefit also extended to the tour guide, who noted that the robot eased the tour conducting by handling navigation and providing additional tour descriptions which the guide alone could not fully manage. On the other hand, concerns arise as blind participants reported distraction from the robot and experiencing heightened self-consciousness in the mixed visual group. 

It is critical to design exhibitions and tours that consider the needs and experiences of both blind visitors and sighted visitors, particularly to mitigate blind visitors' feelings of \textit{``being different.''} The tours should incorporate touch-based exhibits as a core component of the experience and allocate sufficient time for blind visitors to explore objects through touch. Importantly, robotic assistance for blind visitors should not replace or diminish any efforts towards accessible design of exhibitions and tours. Instead, the role of the robot is to reduce barriers that limit blind people's access to these accessible contents. As reported by one participant, they were able to focus more on interacting with the tour guide and the exhibition with the robot's support.  

\revise{It is worth noting that wearable devices such as smartphones and AR glasses~\cite{chang2024worldscribe,gao2025wearable} may also support tour experiences. These devices, while portable, can not guarantee navigation safety for blind people. In terms of experience, the sense of safety and mental ease in navigation are also cited as major benefits for both blind participants and the tour guides, which wearable devices cannot support. Additionally, our robot's sensors cover 270 degrees of its surroundings. In contrast, typical wearable devices only cover the frontal area and would require blind users to turn their heads to receive information on their sides.}



\subsubsection{Multi-stakeholder Interaction}
In answering \textit{\textbf{RQ3:} How would robot-assisted participation affect other members in the group?} Interviews with sighted participants reported relatively slow movement of the robot during the tour, especially in crowded environments, which required them to be more cautious while navigating. Nevertheless, sighted participants reported a positive tour experience, noting that the robot was well integrated into the group and did not affect their enjoyment of the tour. Our tour guide in the field study emphasized the importance of the \textit{Call Guide} feature for ensuring the safety of both the blind visitor and the robot. The tour guide also reported paying extra attention to the robot on the tour, and wished to monitor the robot's status through a separate device, suggesting a need for greater guide engagement in the robot's operation. While mostly following the robot, blind participants also expressed a desire for greater control over the robot, such as adjusting the robot's speed or positioning it closer to the guide. It is a known challenge for the robot to navigate in a crowded environment~\cite{mavrogiannis2023core}, and human intervention can be effective to repair robot failures~\cite{honig2018understanding}. Future work can investigate shared-autonomy control~\cite{kamikubo2025beyond} of the robot in group interactions involving multiple stakeholders and study how blind people perceive and negotiate the robot control shared with the guide or other group members. 


\subsection{Future Work \& Limitation}

\paragraph{Extending Real-World Tour Scenarios.}
Our study tours were conducted in a real-world museum during regular business hours amongst unsuspecting museum visitors. 
However, museums can have a wide range of exhibition types and tour designs can vary significantly in duration, group size, and guide expertise. Our findings are based on small-sized tours in a single science museum, which limits their generalizability to other museum contexts.

The small tour group size (\textit{i.e.}, three visitors and one guide) reflects typical tour sizes at the science museum where our user studies were conducted. Larger tours (\textit{e.g.}, more than 20 visitors) may affect participants' perceptions of the robot's assistive features. For instance, multiple participants imagined that the \textit{Call Guide} feature could be useful in larger groups (see \S~\ref{sec:stakeholder-feedback}). In addition, the science museum encouraged interactions with the exhibits and discussions among tour visitors. Other museums may have different environments, such as a lack of interactive touchable exhibits or a quieter atmosphere (\textit{e.g.}, art galleries~\cite{kreplak2014artworks}). These differences may impact how participants engage with and perceive our robot's assistive features. Lastly, all tours in our study were led by a single museum staff member with training in accessible tours for blind visitors. Other guides may have limited or no knowledge of the accessibility needs of blind visitors. Differences in the tour guide's accessibility knowledge may have introduced bias into our study. Future work should examine robot-assisted group tours for blind people in a broader tour context, including art galleries, gardens~\cite{wang2024translating}, and theaters~\cite{udo2010enhancing}, and investigate how variations in tour size, guide expertise, and environmental context shape group dynamics and the demands of assistive robot features.

\paragraph{Exploring Generalized Group Activities.}

\revise{Although our study focused on small mixed-visual groups, the use of mobile assistive robots could be extended to a wider range of group configurations involving blind people. For example, in a large-group tour designed primarily for blind people, visitors often form a line to touch exhibits one by one. In such contexts, the robot could inform visitors when it is their turn to touch the exhibit and help users avoid touching other blind visitors' hands when multiple people are experiencing the same exhibit. In addition, the robot could manage the flow of the line and help the group maintain an appropriate distance from surrounding crowds to ensure the safety of blind visitors. Further studies in these specific tour contexts are needed to investigate the varying robot design requirements and user interaction patterns.}

In addition, tour groups characterized by intermittent movement and the presence of a guide as the group leader are a subset of the broader group activities spectrum. 
\revise{Extending beyond tours, the robot features developed and investigated in this study have the potential to support a broader range of group-based activities involving blind people. For example, the surrounding description feature could be adjusted to aid blind users in understanding joint attention~\cite{jones2024don, jones2025put}. In dynamic group activities involving movement, such as walking with a group of friends~\cite{repiso2024adaptive}, the robot's group following navigation feature could likewise provide support for blind people for them to focus on social interactions with friends.}


\paragraph{Revisiting Robot Support Features.}
In this paper, we designed a suite of assistive robot features grounded in findings from our interview study and informed by the technical feasibility of implementing the requested features. As tracking technologies continue to advance, several of these robot features can be further enhanced or automated. In the field study, we relied on tele-operation to control the robot's movement and guide-following behavior. We selected tele-operation because following a designated person is still an active research topic~\cite{ye2024person}, especially in crowded museum environments where robust sensor-based human tracking is necessary for the robot to follow the human guide no matter when the human is nearby or at a distance. In the future, we plan to leverage ultra-wideband (UWB) sensors to develop an autonomous guide-following navigation module. We also plan to refine the existing features based on the feedback we received during the field studym including alternative communication with the guide and richer information about nearby people (see \S~\ref{sec:robot-assisted-group-interaction}). Finally, consistent with the findings from \citet{wei2025human}, personalization of robot features is valued by all stakeholders due to diverse individual preferences. Future work should therefore investigate personalization mechanisms for both blind users and tour guides. For example, blind users could have greater control over the robot, such as adjusting the robot's movement speed and the level of detail in surrounding descriptions. Similarly, human guides could tailor the robot's information output, for example, by having the robot give a high-level summary of the exhibition while the guide provides in-depth and personalized explanations.






\section{Conclusion}
Group activities play a central role in social experiences and provide resources and support for achieving shared and individual goals. In this paper, we investigated how assistive robots can support blind people to participate in group activities. Using guided museum tours as a real-world scenario, we designed and implemented a set of assistive robot features to support blind visitors engaging in mixed-visual group tours with a human guide. We conducted an interview study, developed a robotic system informed by these findings, and conducted a field study to understand how blind visitors and stakeholders perceive and interact with the robots. Our findings highlight challenges faced by blind people, including navigating with the group, keeping up with the tour pace, and maintaining connections to the guide. We also demonstrate the potential of assistive robots to address these challenges through features that enhance environmental awareness, facilitate group interaction, and support safe and independent navigation. Future work should focus on strengthening the robot's capabilities by enabling autonomous guide following, providing richer descriptions of the surrounding people and environments, and ensuring customization options for the blind users and the human guide.

\begin{acks}
We would like to thank all participants in our studies. Thank Miraikan for providing the site for our experiments. We would also like to thank Miraikan staff, including Kazuhiko Sugano, Hiromi
Kurokawa and Takashi Suzuki for their help in recruitment, piloting, and running the user studies.  
\end{acks}

\bibliographystyle{ACM-Reference-Format}
\bibliography{reference}

\appendix

\end{document}